\documentclass{aa}  

\usepackage{graphicx}
\usepackage{txfonts}
\usepackage{xcolor}
\begin{document}

   \title{On the incidence rate of RR Lyrae stars with non-radial modes}

   \subtitle{}

   \author{H. Netzel
          \inst{1,2,3}
          \and
    V. Varga \inst{1,4}
          \and
        R. Szab\'o\inst{1,2,4}
          \and
          R. Smolec\inst{5}
          \and
          E. Plachy\inst{1,2,4}
          }

     \institute{Konkoly Observatory, HUN-REN Research Centre for Astronomy and Earth Sciences, MTA Centre of Excellence, Konkoly-Thege Mikl\'os \'ut 15-17, H-1121, Budapest, Hungary
    \and
       MTA CSFK Lend\"ulet Near-Field Cosmology Research Group, H-1121 Konkoly Thege Mikl\'os \'ut 15-17, Budapest, Hungary
     \and
     Institute of Physics, Laboratory of Astrophysics, \'Ecole Polytechnique F\'ed\'erale de Lausanne (EPFL), Observatoire de Sauverny, 1290 Versoix, Switzerland
     \and
    ELTE Eötvös Loránd University, Institute of Physics and Astronomy, 1117, P\'azm\'any P\'eter s\'et\'any 1/A, Budapest, Hungary
    \and
   Nicolaus Copernicus Astronomical Center, Bartycka 18, 00-716 Warsaw, Poland
             }

   \date{Received March 15, 2023; accepted April 16, 2023}

 
  \abstract
   {Over the recent years, additional low-amplitude non-radial modes were detected in many of the first-overtone RR Lyrae stars. These non-radial modes form a characteristic period ratio with the dominant first-overtone mode of around 0.61. The incidence rate of this phenomenon changes from population to population. It is also strongly dependent on the quality of the analyzed data. Current models explaining these additional signals involve non-radial modes of degrees 8 and 9.}
   {Using synthetic horizontal branch populations, we investigate the incidence rate of first-overtone RR Lyrae stars with non-radial modes depending on the population properties, i.e., ages and metallicities. We compare our results with the observed results for globular clusters and the numerous collection of field first-overtone RR Lyrae stars to test the predictions of the models.}
   {We used synthetic horizontal branches combined with pulsation models to predict how the incidence rate would depend on the age and metallicity of the population. To test whether the results based on synthetic horizontal branches are realistic, we compared them to incidence rates observed by TESS in first-overtone field RR Lyrae stars, using photometric metallicity values from a newly established calibration for TESS.}
   {The analysis of synthetic horizontal branches showed that the incidence rate decreases with decreasing metallicity. We inferred photometric metallicity for RR Lyrae stars observed by TESS and showed that the theoretical predictions are in agreement with the observations. Using the same method, we also conclude that the metallicity distribution of RR Lyrae stars showing an additional mode with a period-ratio around $0.68$ appears to be different from that of both all first-overtone stars and those showing additional non-radial modes.}
   {}


   \maketitle
%

\section{Introduction}

RR Lyrae stars are low-mass population II pulsating stars located at the intersection of the horizontal branch (HB) with the classical instability strip. The majority of them pulsate typically in radial fundamental mode (RRab type) or first overtone (RRc type). Double-mode pulsations fundamental mode and first overtone (RRd type) are also often observed among RR Lyrae stars. Double-mode pulsations in the fundamental mode and the second overtone are also known but are rare \citep[see e.g.][]{benko2010}. Six RR Lyrae stars were reported as possible triple-mode pulsators in radial fundamental mode, first overtone and second overtone \citep{jurcsik2015,soszynski2019}.

Besides the dominant pulsations in radial modes, many more phenomena are currently known to occur in RR Lyrae stars. The long-standing mystery is the Blazhko effect, which is a quasi-periodic modulation of the amplitude and/or phase of pulsations. It was discovered more than a hundred years ago by \cite{blazhko}. The Blazhko effect is more common among RRab stars \citep[around 50 per cent, e.g.][]{jurcsik2014} than among RRc stars \citep[around 10 per cent, see][]{netzel_blazhko}. The modulation in RRd stars was also reported, typically in anomalous RRd stars \citep[aRRd, see e.g.][and references therein]{soszynski2016}.

Another phenomenon observed in RR Lyrae stars is the presence of additional signals that cannot correspond to radial modes. Multiple groups of stars with such additional signals were reported already \citep[for a review see][]{mojeReview}. The so-called RR$_{0.61}$ stars are the most common group. These are originally classified as either RRc or RRd stars in which the additional low-amplitude short-period signal forms a period ratio of around 0.60--0.64 with the first overtone. A characteristic feature of this group is that these stars form three sequences in the Petersen diagram, i.e., a diagram of period ratio vs longer period \citep[see e.g. fig. 1 in][]{smolec2017_pet_rev}. The three sequences are at the period ratio of around 0.61, 0.62, and 0.63. Henceforth, throughout this paper, we will use $f_{61},f_{62},f_{63}$ while describing individual signals falling into one of these sequences. Currently, more than a thousand RR$_{0.61}$ stars are known, the majority of which were detected with the ground-based Optical Gravitational Lensing Experiment \citep{netzel_census}, but also numerous sample was found only recently in the space-based missions TESS \citep{benko2023} and K2 \citep{netzel2023_k2}. Additionally, many of the RR$_{0.61}$ stars were detected during studies of smaller samples \citep[e.g.][and references therein]{jurcsik2015,smolec2017,moskalik2015,forro2022,molnar2023}. The explanation of these additional signals was proposed by \cite{dziembowski2016}, who suggested that they are caused by harmonics of non-radial modes of degrees $\ell=8,9$, where harmonic of the non-radial mode of degree 8 corresponds to the $f_{63}$ signals, and of degree 9 corresponds to the $f_{61}$ signals. The $f_{62}$ is a linear combination of the two non-radial modes. We note that similar signals to those in RR$_{0.61}$ stars were detected in classical Cepheids as well \citep[e.g.][and references therein]{soszynski2008,smolec.sniegowska2016,suveges.anderson2018,rajeev,smolec2023}. The model by \cite{dziembowski2016} predicts that the additional signals in classical Cepheids are caused by harmonics of non-radial modes of degrees $\ell=7,8,9$. Interestingly, a similar signal was detected in one anomalous Cepheid only recently by \cite{plachy2021}.

Another interesting multi-mode group of RR Lyrae stars is the so-called RR$_{0.68}$ \citep{netzel068}. These are RRc in which the additional signal, $f_{68}$, has a period longer than the first-overtone and forms a period ratio of around 0.686. Since the first detection more and more RR$_{0.68}$ stars are identified in various stellar systems \citep[e.g.][]{netzel_census,molnar2022,benko2023}. Again, the analogous multi-mode group was also identified among classical Cepheids \citep{suveges.anderson2018,smolec2023}. The only explanation for $f_{68}$ signals was proposed by \cite{dziembowski2016}, but it faces several difficulties (see a discussion in Sec.~\ref{Sec.discussion}). Only recently, \cite{benko_068} reported discovery of analogous signals in RRab stars. 

The incidence rates of RR$_{0.61}$ stars are different depending on the studied sample. The dominant factor influencing the incidence rates is the quality of photometry. \cite{netzel_census} analyzed RRc stars based on the OGLE-IV data for the Galactic bulge fields and obtained an incidence rate of 8.3 per cent for RR$_{0.61}$ stars. On the other hand, \cite{netzel_2fields} analyzed only the two most frequently sampled Galactic bulge fields from the OGLE-IV data and obtained a significantly higher incidence rate of 27 per cent. As expected, using space-based data typically results in a higher incidence rate than using ground-based photometry. From the ground-based data, the incidence rate does not exceed 60 per cent, except for the NGC~6362 study, where the incidence rate of the RR$_{0.61}$ stars is as high as 62.5 per cent of the RRc sample \citep{smolec2017}. From the space-based data, the incidence rate usually exceeds 60 per cent. In this regard, a record holder is the stars from the original {\it Kepler} field, where the inferred incidence rate reaches 100 per cent \citep{moskalik2015,forro2022}. The literature incidence rates of the $RR_{0.61}$ stars from different stellar systems and based on various studies are collected in Table~\ref{tab:incidence_rates}. An interesting result from color-magnitude diagrams is that RR$_{0.61}$ stars tend to avoid the bluest region of it \citep{jurcsik2015,smolec2017,molnar2022}. \cite{netzel.smolec2022} calculated a grid of theoretical models of RR Lyrae stars pulsating in radial and non-radial modes of degrees predicted by \cite{dziembowski2016} and showed that close to the blue edge of the instability strip, these non-radial modes are linearly stable. Consequently, in stellar systems with densely populated horizontal branches, the incidence rate of RR$_{0.61}$ is expected to be lower than 100 per cent due to the lack of excitation of those non-radial modes in the stars close to the blue edge of the instability strip.

Incidence rate is not the only difference between RR$_{0.61}$ stars from different studies. There are also differences in the Petersen diagram between different samples of RR$_{0.61}$ stars \citep[see a discussion in][and their fig. 13]{molnar2022}. The scatter, appearance of characteristic sequences, and average period ratio change from sample to sample. At least some differences between the different samples of the RR$_{0.61}$ stars can be explained by differences in physical parameters, metallicities, and ages. The dependency of radial modes content on those parameters has already been studied from both observational and theoretical perspectives. The occurrence of double-mode RR Lyrae stars in different stellar systems was studied recently by \cite{braga2022}, who reported that the fraction of RRd stars to all RR Lyrae stars increases with decreasing metallicity of the system. \cite{zhang2022} reported that the ratio of fundamental to first-overtone classical Cepheids increases with increasing average population metallicity. The mode selection was studied from a theoretical perspective by \cite{szabo2004} in the context of double-mode RRd stars (see, however, a discussion in \citealt{smolec.moskalik2008revisited} on problems with modeling double-mode RR Lyrae stars).

In this work, we aim to characterize the incidence rate of RR$_{0.61}$ stars, assuming that the signals in these stars are indeed explained by non-radial modes of degrees 8 and 9. First, we calculated synthetic horizontal branches using horizontal branch evolutionary tracks. Then, we used pulsation models to determine how many stars in the horizontal branch would be RRc stars and how many of those would also have non-radial modes of degree 8 or 9 unstable. We varied age and metallicity when calculating synthetic horizontal branches. Based on those results, we can predict the incidence rate of RR$_{0.61}$ stars for different metallicities and ages. We tested our results by comparing them with the numerous sample of field RRc stars observed within the TESS project and analyzed by \cite{benko2023} and to a few globular clusters which were already analyzed in the literature and have derived observed incidence rates of the RR$_{0.61}$ stars. 

Additionally, we present a distribution of photometric metallicity for RR$_{0.68}$ stars based on the numerous sample identified by \cite{benko2023} in the TESS data.

The paper is structured as follows. Methods are presented in Sec.~\ref{Sec:methods}. The results are described in Sec.~\ref{Sec:results} and discussed in Sec.~\ref{Sec.discussion}. Sec.~\ref{Sec:conclusions} summarizes our findings.

\begin{table*}[]
    \centering
    \begin{tabular}{llll}
       System  &  Incidence rate & N$_{{RR}_{0.61}}$ & Reference \\
       \hline
       Galactic bulge (OGLE-III) & 5.6\% & 145 & \cite{netzel1} \\
       Galactic bulge (2 fields of OGLE-IV) & 27\% & 131 & \cite{netzel_2fields} \\
       Galactic bulge (OGLE-IV) & 8.3\% & 949 & \cite{netzel_census} \\
       Pisces K2 & 75\% & 3 & \cite{molnar2015} \\
       TESS (first-light results) & 65\% & 20 & \cite{molnar2022} \\
       TESS & 72\% & 456 & \cite{benko2023} \\
       {\it Kepler} field & 100\% & 4 & \cite{moskalik2015} \\
       {\it Kepler} field ({\it Kepler} Pixel Project) & 100\% & 6 & \cite{forro2022} \\
       
       K2 & 62\% & 452 & \cite{netzel2023_k2} \\
       M80 (K2) & 62.5\% & 5 & \cite{molnar2023} \\
        NGC 5897 (K2) & 11\% & 1 & Kalup et al. (in prep.) \\ 
       M3 & 38\% & 14 & \cite{jurcsik2015} \\
       NGC 6362 & 63\% & 10 & \cite{smolec2017} \\
         
   &
    \end{tabular}
    \caption{Incidence rates of $RR_{0.61}$ stars from the literature. Consecutive columns provide the stellar system (and the particular study), derived incidence rate, the total number of identified $RR_{0.61}$ stars and the reference for the study.}
    \label{tab:incidence_rates}
\end{table*}

\section{Methods}\label{Sec:methods}

\subsection{Synthetic horizontal branches}\label{Subsec:syntheticHB}

We followed the method from \cite{lee1990} to create synthetic horizontal branches for a given age and metallicity. The masses are drawn randomly from the truncated Gaussian distribution:

\begin{equation}
P(M) = \left[ M-(\left\langle M_{\rm HB} \right\rangle - \Delta M) \right]\left( M_{\rm RG}- M \right) \exp\left[ - \frac{\left( \left\langle M_{\rm HB} \right\rangle - M \right)^2}{\sigma_{\rm M}^2} \right],
\end{equation}
where $M$ is the mass of each horizontal branch star, $\left\langle M_{\rm HB} \right\rangle$ is a mean mass of stars on the horizontal branch, $M_{\rm RG}$ is a mass of a star on the tip of the red giant branch if no mass loss occurred, $\Delta M$ is mass lost on average, i.e. $\Delta M = M_{\rm RG} - \left\langle M_{\rm HB} \right\rangle$, and $\sigma_{\rm M}$ is a mass dispersion. We assumed that stars arrive at the horizontal branch at a constant rate regardless of mass, i.e. the age is drawn randomly from a uniform distribution. We chose four values of $M_{\rm RG}$: 0.80\,M$_\odot$, 0.83\,M$_\odot$, 0.85\,M$_\odot$, and 0.88\,M$_\odot$. The $M_{\rm RG}$ value translates to age. That is, the higher the mass, $M_{\rm RG}$, the younger the system. For $Z \approx 0.001$ these masses correspond to the ages of around 13.7, 12, 11, and 9.7\,Gyr, respectively. For lower metallicities, $Z \approx 0.0003$, thses masses correspond to the ages of around 12.5, 11, 10, and 9\,Gyr.
 
We used solar-scaled horizontal branch tracks from the BaSTI database \citep{pietrinferni2004}. From the available tracks, we chose four sets of tracks for the following metal and helium abundances ($Z$, $Y$): (0.0001, 0.245), (0.0003, 0.245), (0.0006, 0.246), and (0.001, 0.246). These values were chosen arbitrarily to cover the metallicity range of RR Lyrae stars and correspond to [M/H]=$-2.267$, [M/H]$=-1.790$, [M/H]$=-1.488$, and [M/H]$=-1.266$, respectively. For each value of mass, $M_{\rm RG}$, and metallicity, we interpolated between the evolutionary tracks using linear interpolation to obtain the synthetic horizontal branch population. Each horizontal branch was populated with 500 stars. We adopted $\sigma_{\rm M}=0.02$ to be constant. We also set $\Delta M=0.15$. We note that in the literature both values are varied. Average mass loss can also be connected to metallicity. In this analysis, however, we did not study those nuances. The principal goal is to study main trends, not to reproduce a particular globular cluster exactly. We also repeated the same analysis using $\alpha$--enhanced tracks from the BaSTI database and the results do not change the conclusion drawn from the analysis of solar-scaled tracks.

\begin{figure}
    \centering
    \includegraphics[width=\columnwidth]{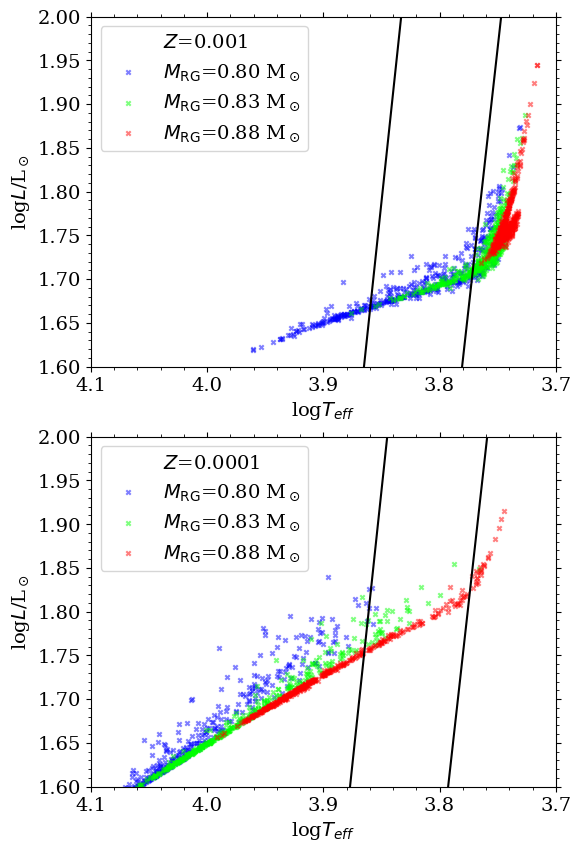}
    \caption{Synthetic horizontal branch stars on the Hertzsprung--Russell diagram. The mass at the tip of the RGB is color-coded as indicated in the key. Black lines are instability strip edges from \protect\cite{marconi2015}. Top panel: for $Z=0.001$. Bottom panel: for $Z=0.0001$.}
    \label{fig:synthetic_age}
\end{figure}

In Fig.~\ref{fig:synthetic_age} we plotted simulated horizontal branch stars for different $M_{\rm RG}$ and two values of metal abundance: $Z=0.001$ (top panel) and $Z=0.0001$ (bottom panel). The population is shifting towards lower effective temperatures for higher $M_{\rm RG}$. For the same $M_{\rm RG}$, the population is shifted towards lower temperatures for higher $Z$. In the case of $Z=0.001$, the population corresponding to $M_{\rm RG}=0.88M_\odot$ is outside the instability strip. In the case of $Z=0.0001$, the population corresponding to the lowest considered mass, $M_{\rm RG}=0.80M_\odot$, is located outside of the instability strip. 

\begin{figure}
    \centering
    \includegraphics[width=\columnwidth]{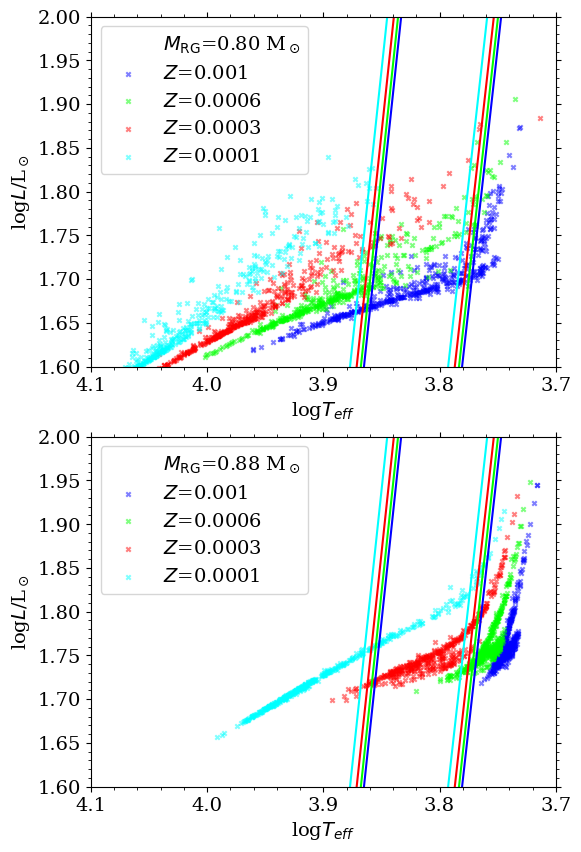}
    \caption{Synthetic horizontal branch stars on the Hertzsprung--Russell diagram. Metal abundance $Z$ is color-coded as indicated in the key. Instability strip edges from \protect\cite{marconi2015} are plotted with lines whose color also corresponds to $Z$. Top panel: for $M_{\rm RG}=0.80$\,M$_\odot$. Bottom panel: for $M_{\rm RG}=0.88$\,M$_\odot$.}
    \label{fig:synthetic_z}
\end{figure}

In Fig.~\ref{fig:synthetic_z} we plotted synthetic HB stars for different metal abundances $Z$, and two different $M_{\rm RG}$: 0.80 (top panel) and 0.88 M$_\odot$ (bottom panel). First, for lower mass, $M_{\rm RG}=0.8M_\odot$, the scatter on the HR diagram is larger. The effect of metal abundance is clearly visible. Namely, the higher the metallicity, the more towards the lower effective temperature is the population shifted. Consequently, inside the instability strip, the stars are shifted towards lower luminosities with increasing metallicity forming a known absolute brightness -- metallicity relation, $M_{\rm V}-$[Fe/H] \citep[e.g.][]{fernley1998}. As already visible from Fig.~\ref{fig:synthetic_age}, for the lowest mass and lowest metallicity, the population is located outside the instability strip. The same situation is for the highest mass and the highest metal abundance. 

We note that in Figures~\ref{fig:synthetic_age} and \ref{fig:synthetic_z} the instability strip edges are plotted based on \cite{marconi2015}, who provides a position of the fundamental red edge and first-overtone blue edge. However, to select first-overtone RR Lyrae stars from the synthetic populations, it is necessary to know the position of the first-overtone red edge. Therefore, we calculated blue and red edge for the first-overtone pulsations using envelope pulsation code MESA--RSP \citep{mesa5}. We calculated a grid of models for each value of metal and helium abundance matching the values of the evolutionary tracks. For each grid we chose one value of mass based on mass--metallicity relation from \cite{castellani1991}. For $Z=0.0001$ we set $M=0.74$\,M$_\odot$, for $Z=0.0003$ we set $M=0.70$\,M$_\odot$, for $Z=0.0006$ we set $M=0.67$\,M$_\odot$, and for $Z=0.001$ we set $M=0.65$\,M$_\odot$. We used $\alpha_{MLT}=1.5$ and convective parameters from set A in \cite{mesa5}. The grid was calculated for three values of luminosity $\log{L}/{\rm L_\odot}\in (1.5,1.6,1.7)$, and for effective temperature from $T_{\rm eff}=5700$\,K to 8000\,K with a step of 100\,K. Linear calculations were carried out and based on the linear growth rates the positions of the edges were interpolated for each metallicity.

\subsection{Incidence rate based on synthetic horizontal branches}

In order to estimate the incidence rate of RRc stars with non-radial modes of degrees 8 or 9, we used the grid of models already calculated by \cite{netzel.smolec2022}. The models were calculated using the envelope code by \cite{dziembowski1977} and are discussed in detail in \cite{netzel.smolec2022}. Here we briefly provide the most important information. The mass range covered by the grid is $0.5-0.9\,{\rm M}_\odot$ with a step of $0.01\,{\rm M}_\odot$. The metallicity range is from --3.0 dex to +0.1 dex with a 0.05 dex step. The luminosity, $\log L/$L$_\odot$, range is 1.3 to 1.8 dex with a step of 0.01 dex, and effective temperature, $\log T_{\rm eff}$, range is from 3.75 to 3.91 dex with a step of 0.005 dex. The number of all models in the grid is over two million. For the further study, we selected four subgrids of models with matching $Z$. On average, each subgrid has around 33\,000 models with linearly unstable first-overtone, and around 15\,000 and 20\,000 were non-radial modes of degrees 8 and 9 are unstable, respectively. We note, that the used envelope code cannot calculate the position of the red edge of the first-overtone instability strip due to the frozen-in approximation of convection. We used the calculated edges described in Sec.~\ref{Subsec:syntheticHB}. We also note that the used edges are linear, while more realistic nonlinear red edge for the first-overtone is hotter.

\begin{figure}
    \centering
    \includegraphics[width=\columnwidth]{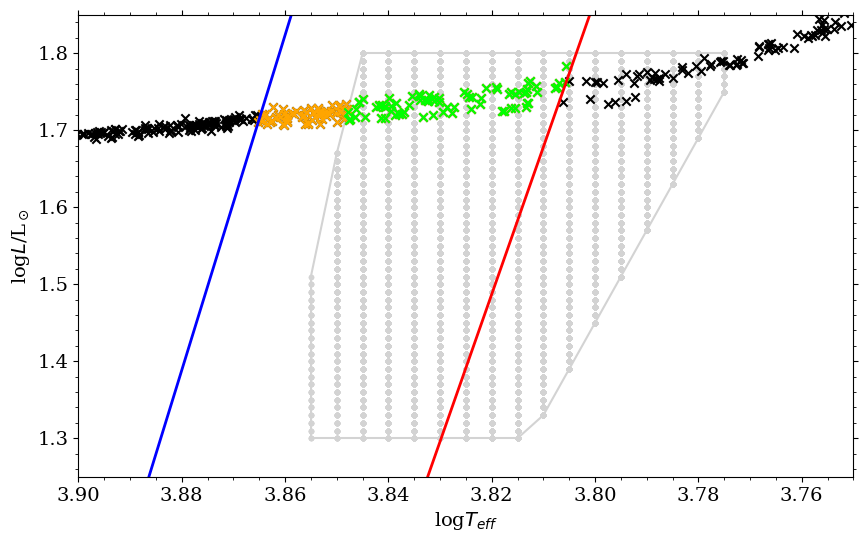}
    \caption{Example of synthetic horizontal branch together with pulsation models used for incidence rate estimation. The models plotted with gray were calculated with the MESA–RSP code by \protect\cite{mesa5}. In these models, at least one non-radial mode is unstable. The red and blue line correspond to the red and blue edge for first-overtone pulsations. Synthetic horizontal branch stars are plotted with black crosses when they are outside of the instability strip, in orange if they have unstable first-overtone but stable non-radial modes, and in green when first overtone and non-radial modes are unstable.}
    \label{fig:example_ir_synth}
\end{figure}

The method to derive the incidence rate based on synthetic horizontal branch and pulsation models is presented in Fig.~\ref{fig:example_ir_synth}. Simulated stars on the synthetic horizontal branch are plotted with crosses. 
The models located in the grey area have at least one non-radial mode linearly unstable. First-overtone red edge is plotted with a red solid line, while blue edge is plotted with a blue solid line. For a given metallicity, we compared the positions of the synthetic horizontal-branch population with selected regions of instability strip provided by theoretical models. In Fig.~\ref{fig:example_ir_synth} the synthetic population is plotted with black if it is outside the instability strip. In green, we plotted those stars that should be RRc stars with additional non-radial modes. Pure RRc stars are plotted with orange. In order to estimate the incidence rate together with its error, we simulated 20 HB populations for a given metallicity and $M_{\rm RG}$. We counted the number of RRc stars with additional non-radial modes and all RRc stars in each simulation and the resulting incidence rate is an average from incidence rates from 20 simulations. The errors were estimated through the bootstrap resampling technique, employing a confidence level of 95 per cent based on percentiles.

\subsection{Metallicity observations}\label{Subsec:photometricFeH}

For testing further if the models trying to explain the extra signals in RRc light curves with high-order non-radial modes are correct, a comparison of the synthetic and the observed metallicity dependencies of incidence rates provides a great opportunity.

For a thorough analysis, good-quality light curves are needed to detect the extra signals in as many cases as possible. In this work, we used TESS photometric measurements processed by \cite{benko2023}. This provides us with 1357 time series analyzed in total for 670 stars brighter than $14~\mathrm{mag}$, from which 23 blended and 14 non-RRc stars were omitted leaving 633 stars for the further analysis. Note that the higher number of time series stems from the fact that some of the stars were observed during multiple TESS sectors.

Spectroscopic metallicities are not available for most of these stars. However, metallicity may be estimated in an empirical way, based on light curve shape and pulsation period. For RRc stars, the formulae were first established by \cite{morgan2007}. 
The hence-derived photometric metallicities are much less precise than those determined by high-resolution spectroscopy, but for a not individual but statistical purpose, as in our case, such approach is sufficient. A difficulty in the photometric metallicity estimations is, however, that the shape of the light curve varies with optical filters, and therefore the calibration of the relation has to be done separately for different passbands. Lacking a fit for TESS, we have established a new calibration for the RRc subtype. Here we present only the key aspects, details are to be found in Varga et al. (in prep.).

As the photometric calibration dataset, the same set as that for the extra signal detection \citep{benko2023} was used, while spectroscopic metallicities were provided by the Carnegie RR Lyrae Survey \citep[CARRS,][]{kollmeier2013,sneden2018}. The cross-matched calibration sample consisted of 96 stars. To describe the light curve, the $P_{\rm 1O}$ first-overtone period and the $\varphi_{31}\equiv \varphi_3-3\varphi_1$ Fourier phase difference were used, along with $\mathrm{[Fe/H]}$ indices describing the metallicity. \citet{benko2023} provide this value for 632 out of 633 stars. For the empirical relation, we ended up using a second-degree polynomial in the following form and getting the following values of the 4 fitting parameters:
\begin{equation}
    \mathrm{[Fe/H]}=-0.26\cdot \varphi_{31}^2-11.5\cdot P_{\rm 1O}+0.96 \cdot  \varphi_{31}+2.05.
\end{equation}
The fitting process did not consider the varying confidence in the data points of the calibration sample, hence the fitting parameters should be considered preliminary, and their final values are yet to be published in Varga et al. (in prep.).
The spectroscopic metallicity values of the calibration sample range from $-2.7$\,dex to $-0.3$\,dex, and the root mean square deviation (RMSD) of the fit is $0.23$\,dex. 
Using such established relation for the \cite{benko2023} TESS sample, an $\mathrm{[Fe/H]}$ value was provided for all 632 stars with known $\varphi_{31}$ values.

Conversion to total $Z$ metallicities is not a straightforward procedure. \cite{kollmeier2013} report that CARRS abundance measurements were performed by fitting to a grid of synthetic spectra generated by MOOG, for which the atomic line lists begun with the \cite{kurucz2011} line database, followed by further refinements. \cite{kurucz2011} uses the solar abundances of \cite{andersgrevesse1989}, who are reportedly \citep{grevesse2013} overestimated. Hence, despite not being the solar reference used for the abundance measurements, we can obtain presumably more accurate $Z$ values calculating with the total solar metallicity value $Z_\odot=0.0134$ of \cite{asplund2009}, approximating the iron contents and total metallicities to be proportional, therefore using the formula
\begin{equation}\label{eq.z}
    Z=Z_\odot\cdot 10^{\mathrm{[Fe/H]}}.
\end{equation}
We note that in case of using a higher solar metallicity reference, like that of \citet{andersgrevesse1989}, the observed trends and therefore conclusions 
remain the same.

\section{Results}\label{Sec:results}

\subsection{Theoretical incidence rate}

\begin{figure}
    \centering
    \includegraphics[width=0.5\textwidth]{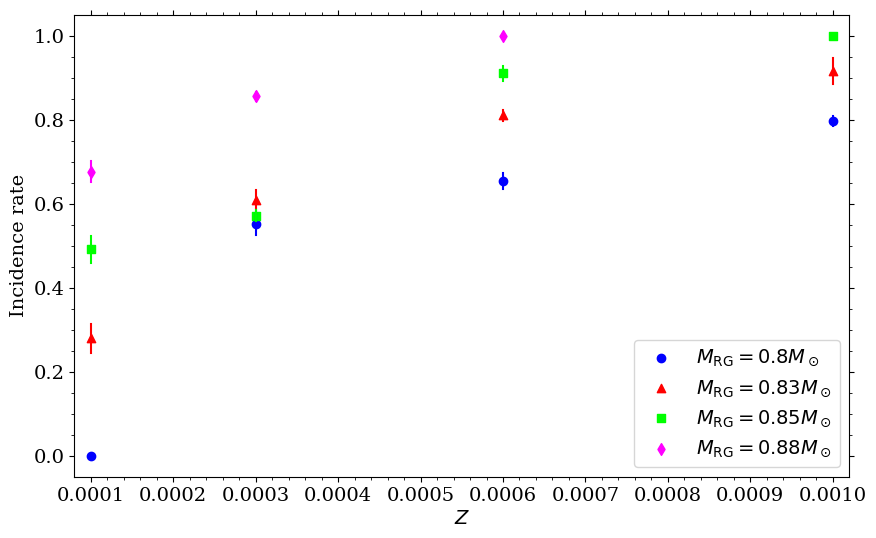}
    \caption{Incidence rate of RR$_{0.61}$ stars as a function of metal abundance, $Z$. The incidence rate is calculated based on synthetic horizontal branches and pulsation models (see text for details). Different colors and symbols correspond to different masses on the tip of the red giant branch as indicated in the key.}
    \label{fig:synth_res}
\end{figure}

Using synthetic horizontal branches for four values of metal abundance, $Z$, and four values of mass at the tip of the red giant branch, $M_{\rm RG}$, we were able to determine the incidence rates for different populations. These results are collected in Table.~\ref{tab:theoretical_ir}, and presented in Fig.~\ref{fig:synth_res}. 
We note, that one population is not plotted. In the case of $M_{\rm RG}=0.88$ M$_\odot$ and $Z=0.001$, i.e. young and metal-rich population, there were no stars inside the instability strip, as the population was redder than the red edge (compare with Fig.~\ref{fig:synthetic_z}).  

There are two trends visible in Fig.~\ref{fig:synth_res}. The incidence rate of the RR$_{0.61}$ stars depends on both metallicity and $M_{\rm RG}$ of the given population. The incidence rate decreases with decreasing metallicity. Moreover, typically, the incidence rate is lower for lower values of $M_{\rm RG}$, which corresponds to older populations. The lowest (non-zero) values of incidence rate are reached by the population of $Z=0.0001$ and $M_{\rm RG}=0.83$\,M$_\odot$, and the highest is for $Z=0.001$ and $M_{\rm RG}=0.85$\,M$_\odot$, and for $Z=0.0006$ and $M_{\rm RG}=0.88$\,M$_\odot$.

\begin{table}[]
    \centering
    \begin{tabular}{llll}
     
       $Z$  & IR & IR$_{\rm min}$  & IR$_{\rm max}$ \\
       \hline
       0.80 M$_\odot$ & & & \\
       \hline
0.001	&	0.79664	&	0.78409	&	0.80882	\\
0.0006	&	0.65485	&	0.63453	&	0.67332	\\
0.0003	&	0.5525	&	0.52515	&	0.58045	\\
0.0001	&	0	&	0	&	0	\\
							
\hline							
0.83 M$_\odot$ & & &  \\							
\hline							
							
0.001	&	0.91642	&	0.88579	&	0.94607	\\
0.0006	&	0.81013	&	0.79762	&	0.8226	\\
0.0003	&	0.6091	&	0.58558	&	0.63313	\\
0.0001	&	0.28059	&	0.24483	&	0.31418	\\
							
\hline							
0.85 M$_\odot$ & & &  \\							
\hline							
0.001	&	1	&	1	&	1	\\
0.0006	&	0.9104	&	0.89128	&	0.92789	\\
0.0003	&	0.57177	&	0.55852	&	0.58548	\\
0.0001	&	0.49106	&	0.45882	&	0.52212	\\
							
\hline							
0.88 M$_\odot$ & & &  \\							
\hline							
0.001	&	--	&	--	&	--	\\
0.0006	&	1	&	1	&	1	\\
0.0003	&	0.85568	&	0.84479	&	0.86656	\\
0.0001	&	0.67526	&	0.65092	&	0.70054	\\
&
    \end{tabular}
    \caption{Theoretical incidence rates. For four different $M_{\rm RG}$ and $Z$, the columns provide the incidence rate (IR) from 20 simulations together with minimum and maximum values of IR corresponding to confidence level of 95 per cent.}
    \label{tab:theoretical_ir}
\end{table}

\subsection{Metallicity of RR Lyrae stars with additional signals}

\begin{figure}
    \centering
    \includegraphics[width=0.5\textwidth]{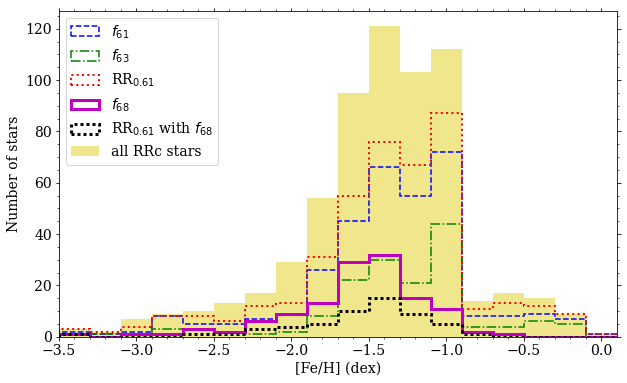}
    \caption{Distribution of photometric metallicity for RRc stars observed by TESS and analyzed by \protect\cite{benko2023}. Yellow bins correspond to all RRc stars. With different line colors and types, we plotted stars that have additional signals detected as indicated in the key. The $f_{61}$, $f_{63}$ and $f_{68}$ refer to the respective signals, while RR$_{0.61}$ represents all RRc stars that show any signal that suggests the presence of the non-radial modes ($l=8$ or $l=9$). This means the inclusion of $f_{63}$, $f_{61}$, $f_{62}$, and their sub-harmonics and combinations frequencies. Note that metallicity values below $-2.7$\,dex or above $-0.3$\,dex are results of extrapolation, and therefore should be treated with caution.}
    \label{fig:obs_feh_inc_dependency_068}
\end{figure}

We determined photometric metallicity for RRc stars studied by \cite{benko2023}. The distribution of metallicities is plotted in Fig.~\ref{fig:obs_feh_inc_dependency_068} for all RRc stars, and for RRc stars that were identified by \cite{benko2023} to show additional signals as indicated in the key. 

The majority of RRc stars in Fig.~\ref{fig:obs_feh_inc_dependency_068} have metallicity from $-2.0$\,dex to $-1.0$\,dex. However, there are stars that have metallicity as low as $-3.5$\,dex and as high as 0.0\,dex. Stars classified as RR$_{0.61}$ or with $f_{61}$ or $f_{63}$ signals have metallicity distributions that do not show large-scale differences compared to the distribution of all RRc stars. Interestingly, stars with $f_{68}$ have a distribution that differs from the distribution of all RRc stars. Typically the RR$_{0.68}$ stars have lower metallicities than all RRc stars, and RRc stars with $f_{61}$ or $f_{63}$ signals. The majority of the RR$_{0.68}$ have metallicity from $-1.7$\,dex to $-1.3$\,dex. Moreover, while RRc stars with and without $f_{61}$ signals can reach relatively high metallicities, the RR$_{0.68}$ are not detected for metallicity above $-0.5$\,dex. The particularly interesting stars are those classified at the same time as RR$_{0.61}$ and RR$_{0.68}$ stars, which are also included in the distribution in Fig.~\ref{fig:obs_feh_inc_dependency_068}. Their metallicity distribution shows more resemblance to the one for stars with $f_{68}$ than to RR$_{0.61}$.

\subsection{Observed metallicity dependence of incidence rates}

\begin{figure}
    \centering
    \includegraphics[width=0.5\textwidth]{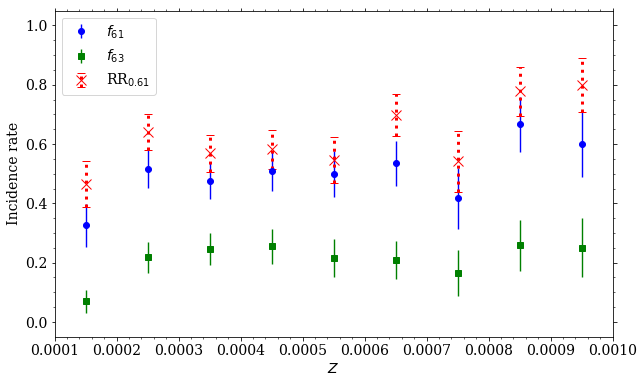}
    \caption{Incidence rates of the $f_{61}$ and $f_{63}$ frequencies and of all RR$_{0.61}$ stars observed by TESS, as function of photometric metal abundance $Z$. The meaning of symbols in the key is the same as in Fig.~\ref{fig:obs_feh_inc_dependency_068}.}
    \label{fig:obs_z_inc_dependency}
\end{figure}

In Fig.~\ref{fig:obs_z_inc_dependency} the incidence rates of RR$_{0.61}$ stars out of all RRc stars are shown, subdivided into groups defined by their more precise period-ratio, namely the $f_{61}$ and the $f_{63}$ signals. We converted the photometrically estimated $\mathrm{[Fe/H]}$ values to total $Z$ metallicities, then calculated the incidence rates using bins from $Z=10^{-4}$ to $Z=10\cdot 10^{-4}$ with a width of $\Delta Z=10^{-4}$. The uncertainties of the incidence rates were calculated as the standard error of the sample mean.

Comparing the observed incidence rates of at least one signal (red crosses)
on Fig.~\ref{fig:obs_z_inc_dependency} to the simulated results on Fig.~\ref{fig:synth_res}, suggests a resemblance in trends. Notably, both distributions indicate lower incidence rates with decreasing metallicity. However, this observation should be verified using metallicity measurements that have better accuracy than those obtained with photometric metallicity calibrations.


\section{Discussion}\label{Sec.discussion}

\begin{figure}
    \centering
    \includegraphics[width=0.5\textwidth]{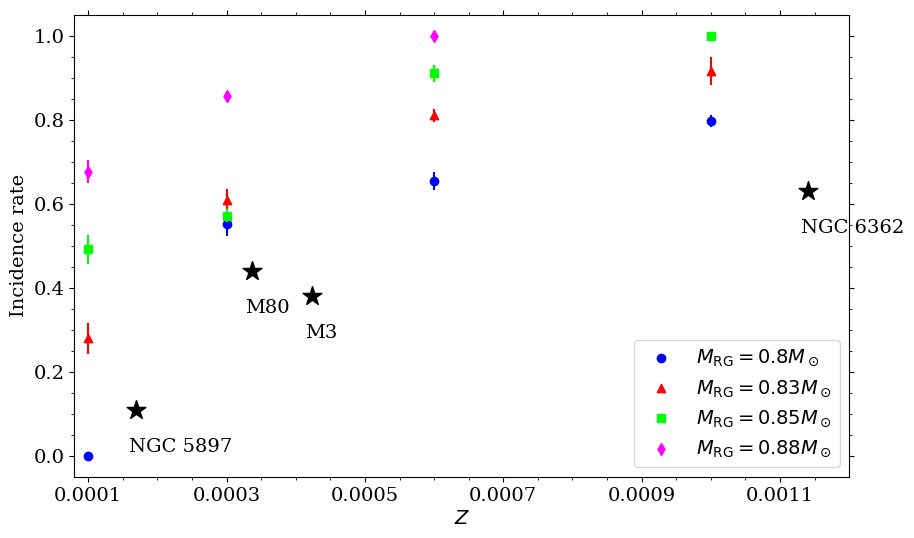}
    \caption{The same as Fig.~\ref{fig:synth_res}, but with four globular clusters added. They are plotted with black star symbols and their IDs are indicated in the figure.}
    \label{fig:gc}
\end{figure}

The predicted trend of decreasing incidence rate of the RR$_{0.61}$ stars with decreasing metal abundance was compared with observed incidence rates for field stars observed with TESS in Fig.~\ref{fig:obs_z_inc_dependency}. The observed and predicted trends, indicating a decrease in incidence rates with decreasing $Z$, seem to align well, suggesting that the simulated results and the underlying theoretical background are justified. The observed values of the incidence rates corresponding to all manifestations of non-radial modes, i.e. $RR_{0.61}$ in Fig.~\ref{fig:obs_z_inc_dependency}, are within the range of the incidence rates based on the models. We note, the uncertainties of photometric metallicities are higher than of those determined spectroscopically. While the trend in observed photometric metallicities from the TESS sample seem to match simulated trends, this conclusion should be verified using metallicities that have smaller uncertainties.

The most uniform populations that we could compare our results to are globular clusters. Several known globular clusters were already analyzed for the occurrence of additional signals in RR Lyrae stars (see Table~\ref{tab:incidence_rates}): M80, NGC\,5897, M3, and NGC\,6362. Interestingly, in the case of NGC\,5897, which is the lowest metallicity of the above-discussed globular clusters, the incidence rate is only 11 per cent (Kalup et al. in prep.). We included these four globular clusters in Fig.~\ref{fig:gc}. For M3 we adopted $\mathrm{[Fe/H]}=-1.50$, for M4 -- $\mathrm{[Fe/H]}=-1.18$, for NGC\,6362 -- $\mathrm{[Fe/H]}=-1.07$, and for NGC\,5897 -- $\mathrm{[Fe/H]}=-1.9$ \citep{carretta2009}. We used Eq.~\ref{eq.z} to calculate $Z$. Interestingly, incidence rates for these four globular clusters appear to follow the same trend with metallicity, although the values of incidence rates are lower than the trend set by the models.

The caveat of the comparison between globular clusters is that they were studied using different instruments and photometric systems. M80 and NGC\,5897 were observed during the K2 mission, while the observations of M3 and NGC\,6362 were carried out in $V$ band using different ground-based telescopes. These differences cause different detection limits of the $f_{0.61}$ signals and the incident rates cannot be directly compared. To estimate the impact of this problem on our analysis we consider the lowest amplitude of an additional signal detected with a given instrument as a rough estimate of its detection limit. This gives a detection limit of 0.5~mmag for M80 and NGC\,5897, 5~mmag for M3, and 2.3~mmag for NGC\,6362. The amplitudes of the $f_{0.61}$ signals are lower for $V$ band than for $I$ band by roughly a factor of two \citep[see fig. 8 in][]{jurcsik2015}. Since the $K_p$ band spans approximately from $I$ to $V$ band, we assume that the amplitude of signals in $K_p$ band is at most twice as large as in the $V$ band. Even with this assumption some stars discovered in K2 would not be detected in the case of M3 and NGC\,6362. Hence, the incidence rate for M3 and NGC\,6362 is likely underestimated. We conclude that indeed the underlying incidence rate for M3 and NGC\,6362 is likely higher than for NGC\,5897 and M80.

The incidence rates observed for globular clusters seem to be lower than based on the models. There are multiple factors that might cause this difference. First, there are observational limitations. In some stars, the additional signals might have amplitudes lower than the noise level and therefore not be identified as RR$_{0.61}$. This is certainly an important factor since the observed incidence rates are strongly correlated with the quality of photometry. Moreover, distributions of amplitudes of additional signals in RR$_{0.61}$ stars show a rise in the number of stars towards low amplitudes until the cutoff due to data limitations (see e.g. figures 4 and 5 in \citealt{netzel_census}, and fig.~11 in \citealt{benko2023}). 

Second, the signals in RR$_{0.61}$ stars are not stable. Specifically, their amplitude and/or phase often vary in time. As a consequence, they form wide structures in frequency spectra, which is commonly observed \citep[see e.g. fig.~9 in][]{netzel2023_k2}. With longer and more precise data it is possible to trace this variability in time. For instance, \cite{netzel_2fields} showed that frequency spectra of RR$_{0.61}$ stars can drastically change from observing season to observing season (see their fig.~10). Long-term monitoring of four RRc stars in the {\it Kepler} field allowed to trace these changes in more detail (see figures 6 and 7 in \citealt{moskalik2015}), which turned out to be significant. The irregularity of changes was also shown based on observations of stars in the continuous viewing zone of TESS (see figures 17 and 18 in \citealt{benko2023}). These changes might contribute to another mechanism that decreases the observed incidence rate. Namely, some RR$_{0.61}$ stars might be simply observed during phases of low amplitude of $f_{61,62,63}$ due to its non-stationary nature, and therefore not identified as RR$_{0.61}$.

The third effect that leads to decrease of the predicted incidence rate is the fact that the non-linear first-overtone red edge is hotter than the linear red edge. Therefore the number of RRc stars with unstable non-radial modes will decrease and consequently the incidence rate will be lower.

As visible from Table~\ref{tab:theoretical_ir}, in our simulated young and metal-rich populations there are no RR Lyrae stars. We note that there are known RR Lyrae stars with high metallicities \citep[e.g.][]{liu2013}, which have kinematics consistent with young thin disk \citep[e.g.][]{iorio.belokurov2021} However, the ages of individual stars are not precisely known to verify this. Moreover, such young and metal-rich RR Lyrae stars are difficult to obtain through standard formation channel. \cite{bobrick2024} showed that they can be created through binary evolution. To test whether young and metal-rich RR Lyrae stars population exists, we used the collection of RR Lyrae stars in galactic globular clusters prepared by \cite{cruzreyes2024}. For metal-rich globular clusters we searched for ages in the literature and checked their population of RR Lyrae stars. Our search revealed that indeed in this sample there is no young and metal-rich globular cluster with sizeable population of RR Lyrae stars. The only candidate is Terzan\,7, whose metallicity is estimated around [Fe/H]$\approx -0.6$ \citep{harris1996} or [Fe/H]$\approx -0.6$ \citep{sbordone2005,tautvaisiene2004}. Its age is estimated to be 9 Gyr by \cite{buonanno1995} or 6 Gyr by \cite{tautvaisiene2004}. However, this cluster contains only one RR Lyrae star \citep{cruzreyes2024}.

\subsection{RR$_{0.68}$ stars}

The analysis of photometric metallicities for a large sample of RRc stars from the TESS data leads to additional noteworthy results regarding the RR$_{0.68}$ stars. The metallicity of the RR$_{0.68}$ stars have a different distribution than all RRc stars, and the metallicity of the RR$_{0.68}$ stars is typically lower. We applied the Komogorov-Smirnov test to verify the difference between the two distributions. We assumed that the uncertainity of all metallicity measurements is 0.23 dex, i.e. RMSD of the fit (see Sec.~\ref{Subsec:photometricFeH}). The test yielded a Kolmogorov-Smirnov statistic of 0.19 and a p-value of 0.001. With a significance level of 0.05, the null hypothesis that the two samples were drawn from the same distribution was rejected. This indicates a statistically significant difference between the distributions of the RR$_{0.68}$ and all RRc stars.

This is important in the context of explaining the nature of the RR$_{0.68}$ stars. At present, the nature of the additional signal is unknown. \cite{dziembowski2016} showed that the period ratio of around 0.68 can be reproduced by pulsations in radial fundamental mode and first overtone if the RR$_{0.68}$ stars are in fact not true RR Lyrae stars, but low-mass giants stripped from their envelopes similarly to binary evolution pulsator \citep{pietrzynski2012}. This explanation, however, is challenged by stars that show $f_{61,62,63}$ and $f_{68}$ signals simultaneously. Such stars were already detected from the ground and space-based photometry \citep[e.g][]{netzel_census,benko2023,netzel2023_k2}. Therefore explanations proposed for RR$_{0.61}$ stars and for RR$_{0.68}$ stars are mutually exclusive. We note that $f_{68}$ signals were also detected in classical Cepheids since the development of the theory by \cite{dziembowski2016}, and only recently in RRab stars \citep{benko_068}. So far, there was no clear observed difference between the RR$_{0.61}$ and the RR$_{0.68}$ stars. In this study, we show for the first time that in fact, the RR$_{0.61}$ and the RR$_{0.68}$ stars show differences. This observation will hopefully contribute to future efforts in explaining the nature of the additional signals in the RR$_{0.68}$ stars.

\section{Conclusions}\label{Sec:conclusions}

We calculated synthetic horizontal branch populations to investigate the incidence rate of stars with non-radial modes of degrees $\ell=8,9$. We calculated populations for four values of metallicity, ($Z$,$Y$): (0.001, 0.246), (0.0006, 0.246), (0.0003, 0.245), and (0.0001, 0.245), and for four values of mass at the tip of the red giant branch, $M_{\rm RG}$: 0.80\,M$_\odot$, 0.83\,M$_\odot$, 0.85\,M$_\odot$, and 0.88\,M$_\odot$. We used BaSTI horizontal-branch evolutionary tracks \citep{pietrinferni2004}. We used MESA-RSP code by \cite{mesa5} to derive the instability strip edge for first-overtone pulsations, and a grid of models from \cite{netzel.smolec2022} to get the region of the instability strip where the non-radial pulsations in modes of degrees 8, and 9 occur.

We compared our theoretical predictions of incidence rates with observed incidence rates of RR$_{0.61}$ stars. We used four globular clusters with derived incidence rates: NGC 5897 (Kalup et al. in prep.), M80 \citep{molnar2023}, M3 \citep{jurcsik2015}, and NGC 6362 \citep{smolec2017}. We also used RRc stars observed by TESS and analyzed by \cite{benko2023}. For this sample, we derived photometric metallicities based on our newly established calibration for TESS. Our findings are listed below:

\begin{itemize}
    \item For a population with metallicity of $Z=0.001$ and mass at the tip of the red giant branch, M$_{\rm RG}$=0.88\,M$_\odot$, i.e. metal-rich and young population, all stars were outside of the instability strip. In other words, this population did not contain any RR Lyrae stars.
    \item For the rest of the combinations the incidence rate varied from high values, close to 100 per cent, to relatively low (non-zero) values of almost 20 per cent.
    \item The theoretically predicted incidence rate decreases with decreasing metallicity.
    \item The theoretically predicted incidence rate is lower for lower values of M$_{\rm RG}$, i.e. for older populations.
    \item The trend of decreasing incidence rate with decreasing metallicity appear in agreement with the observed values of incidence rates for four globular clusters. The values of observed incidence rates are systematically smaller than predicted by the simulations (see a discussion in \ref{Sec.discussion}). 
    \item The RR$_{0.61}$ stars from the TESS sample also show the slight trend of decreasing incidence rate with decreasing metallicity but are subject to the relatively significant error bars.
\end{itemize}

We investigated differences between metallicity distributions for the whole RRc TESS sample and stars classified as RR$_{0.61}$ or RR$_{0.68}$. The distribution of photometric metallicity for the whole RRc TESS sample and the RR$_{0.61}$ stars do not show large scale differences. Interestingly, RRc stars with the $f_{68}$ signals show a metallicity distribution that significantly differs from the rest of the RRc sample. RR$_{0.68}$ stars tend to have typically lower metallicities. This is also the case for the RR$_{0.61}$ stars with the $f_{68}$ signals.

\begin{acknowledgements}
This work has made use of BaSTI web tools.
This project has been supported by the Lend\"ulet Program of the Hungarian Academy of Sciences, project No. LP2018-7/2022, the `SeismoLab' KKP-137523 \'Elvonal grant and OTKA projects K-129249 and NN-129075 of the Hungarian Research, Development and Innovation Office (NKFIH), as well as the MW-Gaia COST Action (CA18104). This work was supported by the European Research Council (ERC) under the European Union’s Horizon 2020 research and innovation programme (Grant Agreement No. 947660). VV is supported by the ÚNKP-22-1 New National Excellence Program of the Ministry for Culture and Innovation from the source of the National Research, Development and Innovation Fund, and by the undergraduate research assistant program of Konkoly Observatory. RSm acknowledges financial support by the Polish National Science Centre (NCN) under Sonata Bis grant 2018/30/E/ST9/00598. We thank K. I\l kiewicz for constructive discussions.

\end{acknowledgements}

\bibliographystyle{aa} 
\bibliography{references.bib} 

\end{document}